\documentstyle[12pt,epsf]{article}

\newcommand{\br}{{\bf r}}
\newcommand{\bx}{{\bf x}}
\newcommand{\ba}{{\bf a}}

\newcommand{\by}{{\bf y}}
\newcommand{\bB}{{\bf B}}
\newcommand{\bE}{{\bf E}}
\newcommand{\bD}{{\bf D}}
\newcommand{\bG}{{\bf G}}
\newcommand{\bv}{{\bf v}}
\newcommand{\bb}{{\bf b}}

\begin{document}

\font\ninerm = cmr9

\baselineskip 16pt plus .5pt minus .5pt

\def\footnoterule{\kern-3pt \hrule width \hsize \kern2.6pt}

\hsize=6.4truein
\vsize=9.0truein
\textheight 8.5truein
\textwidth 5.7truein
\voffset=-.4in
\hoffset=-.5in

\pagestyle{empty}
\begin{center}
{\large\bf  Dynamics  of 
BPS States in 
the Dirac-Born-Infeld Theory}\footnote{\ninerm
\hsize=6.0truein This work was
supported in part by funds provided by 
the Korea Science and Engineering Foundation through
the SRC program of SNU-CTP,
the Basic Science  Research Program
under project \#BSRI-97-2425/2441, and 
University of Seoul through
the Academic Research Program.}
\end{center}

\vskip 1.5cm

\begin{center}
{\bf Dongsu Bak\footnote{\ninerm
\hsize=6.0truein email address: dsbak@mach.uos.ac.kr},
Joohan Lee, and Hyunsoo Min}
\end{center}
\begin{center}
{\it Department of Physics,
University of Seoul,
Seoul 130-743, Korea}
\end{center}

\vspace{1.2cm}
\begin{center}
{\bf ABSTRACT}
\end{center}
The Dirac-Born-Infeld action with 
transverse scalar fields is considered
to study the dynamics of  various BPS states. 
 We first describe the characteristic properties 
of the so-called 1/2 and 1/4 BPS states on the D3 brane, 
which can be interpreted as
F/D-strings ending on a D3-brane in Type IIB string theory
picture. We then study the response of the BPS states
to  low energy excitations of massless fields on the brane, 
the 
scalar fields representing the shape fluctuation  of the 
brane and   U(1) gauge fields describing the open string
excitations on the D-brane.
 This leads 
to an identification of interactions between
 BPS states including the static potentials and  the 
kinetic interactions.

\bigskip
\begin{center}
(PACS:14.80.Hv, 11.15.Kc, 11.15.-q)
\end{center}
\bigskip
\bigskip
\bigskip

\vfill
\space\space UOSTP-98-102 \space\space
 \space\space
SNUTP-98-059 
\hfill
\eject

\baselineskip=14pt plus 3pt minus 3pt

\pagestyle{plain}
\pagenumbering{arabic}
\setcounter{page}{1}

\pagebreak[3]
\setcounter{equation}{0}
\renewcommand{\theequation}{\arabic{section}.\arabic{equation}}
\section{Introduction}
\nopagebreak
\medskip
\nopagebreak

The Dirac-Born-Infeld theory\cite{born,dirac} has 
recently drawn much attention 
in relation with the world-volume dynamics of 
Dp-branes\cite{callan}-\cite{hashimoto}. 
The degrees of freedom involved with D-branes,   consist of 
 world-volume gauge fields  and a number of scalar fields,
representing  the shape fluctuations 
of the Dp-brane embedded in higher dimensions. 
This may be understood from the fact that 
the $p+1$-dimensional Dirac-Born-Infeld action for Dp-brane 
is the dimensional reduction of the ten-dimensional 
supersymmetric Born-Infeld electromagnetism. The action
allows static BPS configurations that saturate the so-called
Bogomol'nyi bound. As shown in Ref. \cite{callan}, they  represent 
attachments of 
F(fundamental)-strings or D(Dirichlet)-strings to the Dp-brane. 
The dynamics of these 
states on the world-volume will be the main 
concern of this note.
   
To be specific, we shall focus on the case of D3-brane, 
whose $3+1$-dimensional 
world-volume 
is  immersed in ten spacetime so that the  
transverse space is  six dimensions. 
As explained, for example, in  Ref. \cite{callan},
 the 
electrically charged 
BPS states describe the intersection 
of the D3-brane with F-string,
and  the magnetic states describe   
the D-string that ends on the brane. There are also
 dyonic $(q,g)$-strings that may be viewed as  the bound states of 
F-string and D-string. The magnitude of
the  scalar 
charge restricted by the BPS
condition, can be identified with  the tension of the attached string, 
while the direction
of the string in the transverse space 
is specified by the $SO(6)$ direction of the scalar 
charge. Especially, the scalar charge determines how the D3-brane is 
pulled out
by the string\cite{hashimoto}.
 
In this note, we shall first identify possible BPS states by 
investigating the energy functional of the 
Dirac-Born-Infeld theory. These
will include  $1/2$-BPS states such as 
electric-pole, magnetic monopole, and 
dyonic states where the attached strings are 
 directed all in one direction in the 
transverse six space. We shall 
also describe the properties of a $1/4$ BPS state\cite{bergman} 
that represents
a monopole and an electric-pole pulled out in mutually
perpendicular 
directions in the transverse space.
It will be shown that there are  more general $1/4$ BPS 
configurations.
We present explicitly
 these solutions as well as their characteristic properties.  

We then study  
 the responses of 
various BPS states to asymptotic massless excitations by 
exploiting the classical field equations. As will be shown 
below,
the motion of the center positions of F/D-strings will be 
governed by
a generalized Lorentz force law. 
Furthermore, the acceleration will be accompanied
by the electromagnetic and  the scalar radiations.  
We, then, turn to the case 
where
 multiple strings of various charges are joined to the D3-brane with 
 different $SO(6)$ angles.
Of course, they may interact with each other via the electromagnetism 
or  the deformation of the shape of the D3-branes. We shall
first determine the static potentials between the F/D-strings 
 as a function of 
their separation and charges. The 
nature of the scalar interaction will be 
explicitly exposed for various cases.
In particular, when 
the  static potential between the F/D-strings  
vanishes, the  multi-string states can be static;
only kinetic interactions
between the F/D-strings will appear in this case 
when they moving 
around on the parallel three space of the D3-brane.
For  generic non-BPS 
configurations, we shall also find the kinetic 
interactions as well 
as the  static interactions. 
   
Section II is devoted to the various BPS states 
including electric, magnetic 
and dyonic states that break half of the 
supersymmetries. We shall also 
discuss  $1/4$-BPS states that leave four 
supersymmetries unbroken out
of  sixteen supersymmetries.  We shall 
identify all the charges and 
illustrate the shape of the solutions.

In Section III, the response of the BPS states 
to the world-volume massless 
fields will be identified by studying the fields 
equations, which may be 
summarized  
by the force law for the rigid motion of the F/D-strings. 
We will also discuss 
the radiations produced by  acceleration. We also study the 
response of the states to the incident  scalar  waves  or 
electromagnetic waves, and discuss the nature of the 
scattering of these waves
by the BPS configurations.

In Section IV, we will obtain the low-energy 
effective  Lagrangian that 
describes all two body interactions between various states. The static  
interactions together with the kinetic interactions will be exploited 
based on the effective Lagrangian.

 Section V comprizes discussions and conclusions.

\pagebreak[3]
\setcounter{equation}{0}
\renewcommand{\theequation}{\arabic{section}.\arabic{equation}}
\section{Dirac-Born-Infeld Theory and its BPS States}
\nopagebreak
\medskip
\nopagebreak

We begin with the 
Dirac-Born-Infeld Lagrangian 
for a single D3-brane 
\begin{equation} 
L=T\int d^3 x (1- K^{{1\over 2}})  \label{lag} 
\end{equation}
with
\begin{equation} 
K=-{\rm Det}(\eta_{\mu\nu}+\partial_\mu Y^I
\partial_\nu Y^I+ F_{\mu\nu} )  \label{det} 
\end{equation} 
where $F_{\mu\nu}=\partial_\mu A_\nu-\partial_\nu A_\mu$ and 
the index $I$ for the transverse space runs from $1$ to 
$6$\footnote{We will use conventions $\eta_{\mu\nu}=
{\rm diag}(-,+,+,+)$ and $T=1$}.
As is well known, this Lagrangian is 
the dimensional reduction of the ten-dimensional pure Born-Infeld
action,
\begin{equation} 
L=T_{10}\int d^{10} x [1- (-{\rm Det}(\eta_{MN}+
F_{MN} ))^{{1\over 2}}]  \label{lag10} 
\end{equation}
to $3+1$ dimensions where  $A_M$ is identified  
with the 
world-volume gauge potential for $M=0,1,\cdots,3$ and 
with the scalar fields $Y_{M-3}$ for $M=4,5,\cdots, 9$.

Let us first consider the case where the attached string is directed 
in one direction, $\hat{e}^I$ in the transverse space. As will be seen 
below, this includes  the configurations of 
the monopole, electric-pole, and
dyon.
One may here consistently set all the other perpendicular 
 components of $Y^I$ 
to be zero except
$Y^I \hat{e}^I$, which we denote $\phi$. The determinant $K$ 
in  this case can be explicitly written as
\begin{equation} 
K=1-{\dot\phi}^2 - {\bf E}^2 + (\nabla\phi)^2 +{\bf B}^2
+({\bf B}\cdot \nabla\phi)^2 -({\bf B}\cdot {\bf E})^2-
(\bE\times \nabla\phi + {\dot\phi}\bB)^2
\label{detone} 
\end{equation}
where $E_i=F_{i0}$ and $B_i={1\over 2}\epsilon_{ijk}F_{jk}$.
To the quadratic order in its dynamical
variables, this Lagrangian corresponds to the 
conventional electromagnetism
with a free scalar field. This picture will be valid when the involved
field strengths are weak enough. But the nonlinearity in the region of 
strong fields
 will play an
important role to understand the brane dynamics. 
 
As discussed in Appendix in a more general setting, 
the above second-order Lagrangian may be 
turned into a first-order form,   
\begin{equation} 
 {\cal L}=\Pi {\dot \phi}- \bD\cdot {\dot{\bf A}}-A_0 \nabla\cdot \bD
-{\cal H}
\label{firstorder} 
\end{equation}
where the Hamiltonian density (${\cal H} \ge  0$) is given by
\begin{equation} 
 ({\cal H}\!+\!1)^2=1+{\Pi}^2+ {\bf D}^2 + (\nabla\phi)^2 +{\bf B}^2
+( {\bf D}\cdot\nabla\phi)^2+({\bf B}\cdot \nabla\phi)^2 +
(\bD\times \bB + {\Pi}\nabla\phi)^2\,.
\label{hamiltonian} 
\end{equation}
In order to find  the Bogomol'nyi bound\cite{bogomolnyi} 
for the static configurations, 
we rewrite
the Hamiltonian into 
 the form,
\begin{eqnarray} 
 ({\cal H}\!+\!1)^2
&\!\!\!=\!\!\!&{\Pi}^2+ (\bD\times \bB + {\Pi}\nabla\phi)^2
+(\sin\xi\bB\cdot \nabla\phi-\cos\xi\bD\cdot \nabla\phi)^2
+(\bB -\cos\xi \nabla\phi)^2
\nonumber\\ 
&\!\!\!+\!\!\!&(\bD -\sin\xi \nabla\phi)^2 
+( 1+\cos\xi {\bf B}\cdot\nabla\phi + 
\sin\xi{\bf D}\cdot \nabla\phi)^2 \,.
\label{hamiltoniana} 
\end{eqnarray}  
From this expression, one may easily recognize that the 
Hamiltonian density is bounded below by\cite{gauntlett,silva} 
\begin{eqnarray} 
 {\cal H} \ge \cos\xi {\bf B}\cdot\nabla\phi + 
\sin\xi{\bf D}\cdot \nabla\phi\,.
\label{saturation} 
\end{eqnarray}   
The saturation of the bound occurs if the  Bogomol'nyi
equations,
\begin{eqnarray} 
\bB =\cos\xi \nabla\phi, \ \  
\bD =\sin\xi \nabla\phi, \ \ \Pi=0\,,
\label{bogomolnyi} 
\end{eqnarray}   
 together
with the Gauss constraint, $\nabla\cdot\bD=0$, 
are satisfied. This set of 
equations may also be derived from the 
consideration of the supersymmetric variation of the 
gaugino\cite{callan}, from which one finds that the BPS background
satisfying the above equations  preserves only  
half of the sixteen supersymmetries. A static solution of 
the Bogomol'nyi equation is found to be
\begin{eqnarray}  
\bD ={q\hat{r}\over 4\pi r^2}\,,\ \ \bB=
{g\hat{r}\over 4\pi r^2}\,,\ \ 
\phi=-{q_s\over 4\pi r}\,,
\label{solution} 
\end{eqnarray}      
with relations between charges, $g=\cos\xi\, q_s$ and 
$q=\sin\xi\, q_s$. Here we follow the conventional 
definitions 
\begin{eqnarray}  
Q_E =\oint_{r\rightarrow \infty} dS^i D_i\,,\ \ 
Q_M =\oint_{r\rightarrow \infty} dS^i B_i
\label{emcharges} 
\end{eqnarray}      
respectively 
for    the electric and  
the magnetic charges. In addition we   define
the scalar charge by
\begin{eqnarray}  
Q^I_S =\oint_{r\rightarrow \infty} dS^i \partial_i Y^I\,,
\label{scalarcharge} 
\end{eqnarray}
which is a six vector in the  transverse space.


\begin{figure}
\epsfxsize=3.5in
\centerline{
\epsffile{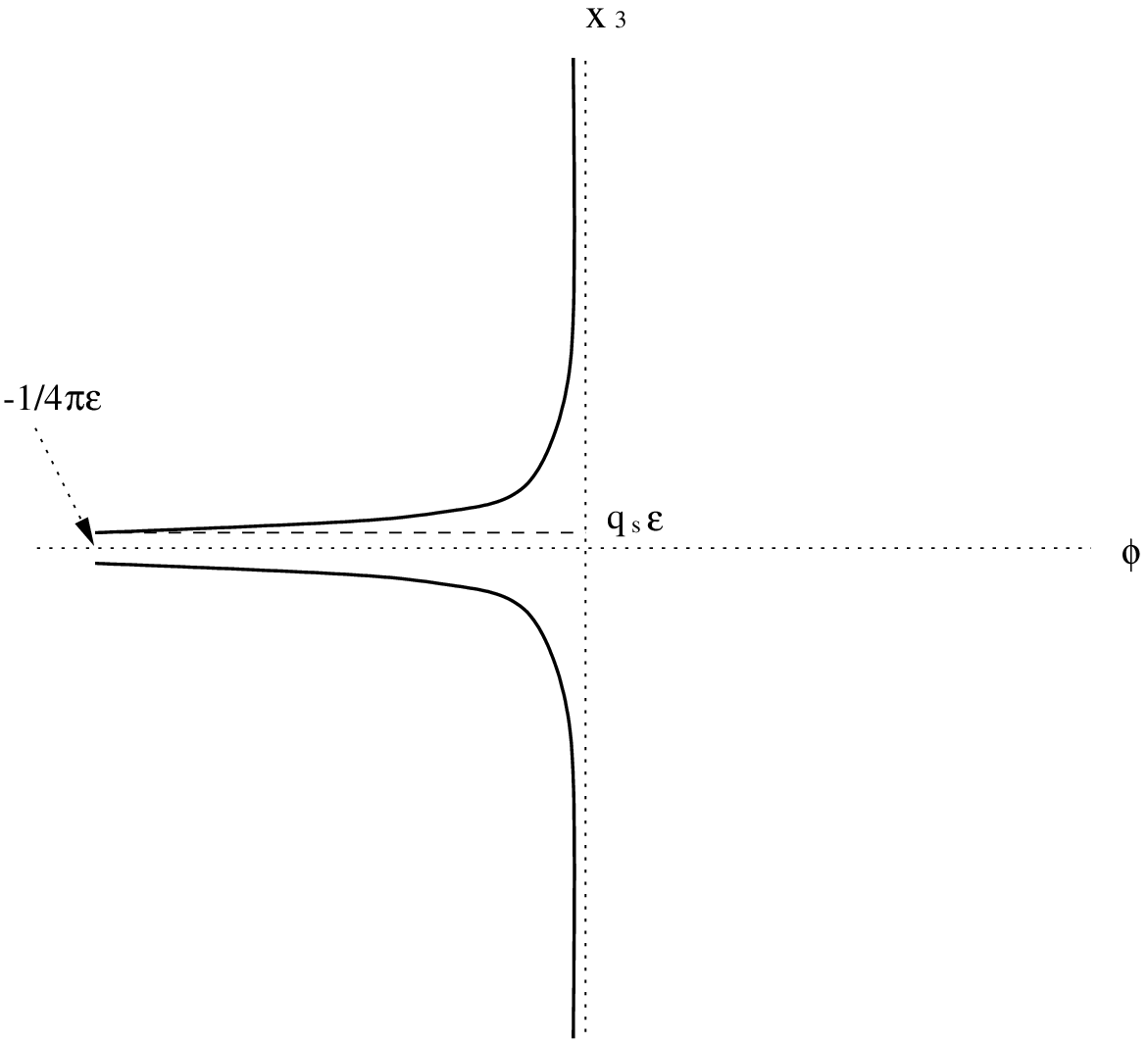}
}
\vspace{.1in}
{\small Figure~1:~The shape of a  D-string located at $r=0$, 
is drawn in $x_3-\phi$ plane.}
\end{figure}


 The shape of this solution seen in $x_3-\phi$ 
plane is depicted in Fig. 1 for $q_s > 0$. The spike in the transverse
direction represents the F/D string attached to the D3 
brane. Namely, the F/D string pulls out the D3 
brane in a smooth manner and the corresponding strength 
given by the scalar charge  is proportional to the tension
of the string.   
The total energy defined by 
the spatial integration of the Hamiltonian density ${\cal H}$
is infinity since the string is stretched to the infinity in the 
transverse space. But we regulate it by restricting the domain of 
the integration by $r\ge |q_s|\epsilon$.
(The physics at $r=|q_s|\epsilon$ will be determined by specifying
consistent  boundary data, whose dynamical aspects  will be 
further discussed in the next Section.) 
 In this case, the regulated mass of the 
configuration is    
\begin{eqnarray}  
M(\epsilon)= |q_s| l(\epsilon)\,,
\label{mass} 
\end{eqnarray}     
with  the transverse length $l(\epsilon)$ of the string being 
$|\phi(\epsilon)-\phi(\infty)|=1 /(4\pi\epsilon)$.
Since the scalar charge serves as the tension of the 
attached F/D-string,
the mass agrees, as expected, with the tension multiplied 
by the length
of the string\cite{callan}.
  
The solution (\ref{solution})  describes a few distinctive
  cases. When the scalar
 charge is   positive, it corresponds to the string stretched 
to the direction $-{\hat{e}}^I$, whereas the negative signature 
of the charge implies that the string is stretched to the 
opposite direction, i.e. ${\hat{e}}^I$. Since the unit vector 
${\hat{e}}^I$ is  in arbitrary  direction in the transverse space,
the F/D-string may then be in any transverse direction; the freedom 
in the signature of the scalar charge reflect the remaining 
possibility when an axis $\phi$ is chosen.  There is also an angle  
parameter 
$\xi$. For  $\xi= \pi/2, 3\pi/2$, the magnetic field vanishes and,
consequently, the string is electrically charged. It is  nothing but 
the fundamental  string ending on the D3-brane.
 On the other hand, the magnetically charged 
string (D-string) corresponds to the case of $\xi=0,\pi$.
The remaining generic value of $\xi$ is for a bound state 
of F-string and D-string, i.e. a dyon. Furthermore, the 
signature of electric 
or magnetic charges carried by the attached string distinguishes whether
one deals  with string or anti-string; in other words, it correspond to 
the orientation of the F/D-string. 

There are more generic BPS solutions than (\ref{solution}), which 
correspond 
to multi-strings located in  various positions on 
the D3-brane. They are given by
\begin{eqnarray}  
\bD =-\sum_n \nabla {q_n\over 4\pi |\br-\bx_n|}\,,\ \ \bB=
-\sum_n \nabla {g_n\over 4\pi |\br-\bx_n|}\,,\ \ 
\phi=-\sum_n {(q_s)_n\over 4\pi |\br-\bx_n|}\,,
\label{multisolution} 
\end{eqnarray}    
with  relations between charges, $g_n=\cos\xi \, (q_s)_n$ and 
$q_n=\sin\xi \, (q_s)_n$ for all $n$. The sum of each dyon mass 
that is given by (\ref{mass}) will be the total 
mass of the configuration.
 Since they are 
 BPS configurations,
the scalar  and the electromagnetic forces between any two  F/D-strings 
are exactly
canceled  for all cases. There are two distinct situations in the two body 
interactions. One is the case where the two F/D-strings are 
stretched in the same 
direction in the transverse space. The signs of the scalar charges
are the same and, hence, the scalar interaction is 
attractive. The cancellation 
of the forces in this case occurs 
since the electromagnetic force is repulsive with 
the same strength. The other 
corresponds to the situation 
where  the spikes of the strings are 
in the exactly opposite directions.
The scalar force  is repulsive in 
this case, and the attractive 
electromagnetic force cancels it.

Having analyzed the aligned F/D-strings in the transverse
space, let us now turn to the case where two transverse 
directions are allowed so that spikes of F/D-string may 
have an $SO(6)$-angle between them. Since a F/D string may have its 
spike into an arbitrary transverse direction specified by a unit 
transverse vector, 
the two unit vectors involved with two F/D strings form
a plane   in general 
in the transverse space.  Let us  denote  the orthonormal
unit vectors in this plane by ${\hat{e}}_1^I$ and  
${\hat{e}}_2^I$. Then one may consistently set all the components 
of the scalar perpendicular to the plane to be zero.
 The nonvanishing component are restricted on 
the plane, which we  denote by 
$Y^I=\phi {\hat{e}}_1^I + \chi{\hat{e}}_2^I$.  
With these two component scalar fields, the determinant in 
(\ref{det}) can be evaluated as
\begin{eqnarray} 
K&=&1 +{\bf B}^2 + (\nabla\phi)^2
+(\nabla\chi)^2
+({\bf B}\cdot \nabla\phi)^2 +({\bf B}\cdot \nabla\chi)^2 
+(\nabla \phi\times \nabla\chi)^2
\nonumber\\
&-&{\dot\phi}^2-{\dot\chi}^2 - {\bf E}^2 -({\bf B}\cdot {\bf E})^2-
(\bE\times \nabla\phi + {\dot\phi}\bB)^2 -
(\bE\times \nabla\chi + {\dot\chi}\bB)^2
\nonumber\\
&-&({\dot\phi\nabla\chi-\dot\chi\nabla\phi})^2
-(\bE \cdot \nabla\phi\times \nabla\chi + \dot\phi\nabla\chi
\cdot \bB -
\dot\chi\nabla\phi
\cdot \bB)^2\,.
\label{dettwo} 
\end{eqnarray}
Although its complicated appearance, 
one may easily recognize that it does possess the $SO(2)$-rotational 
invariance in the transverse plane under the transformation,
$\phi'=\cos\alpha \phi-\sin\alpha \chi$ and 
$\chi'=\sin\alpha \phi+\cos\alpha \chi$.
In a Lorentz invariant form, it may be rewritten as
\begin{eqnarray} 
K=1 + {1\over 2}F^2_{mn} -{1\over 2}\Bigl({1\over 8}\epsilon^{mnpqrs}
F_{pq}F_{rs}\Bigr)^2
-\Bigl({1\over 48}\epsilon^{mnpqrs}F_{mn}F_{pq}F_{rs}\Bigr)^2\,,
\label{detttt}    
\end{eqnarray}
where the indices run from zero to five and $F_{mn}=\partial_m A_n-
\partial_n A_m$ with $(A_4,A_5)=(\phi,\chi)$.
In this Lagrangian,  one may consistently set, for instance, 
$\chi$ to zero, and, then, the system is reduced  to the case of one 
scalar
discussed previously. The Bogomol'nyi equations in 
(\ref{bogomolnyi})  are again serving as  the condition 
for the BPS saturation. There are, of course, a family of Bogomol'nyi
equations that are produced by $SO(2)$ rotation; for example, 
the Bogomol'nyi equation (\ref{bogomolnyi}) 
with $\chi$ instead of $\phi$ is obtained by the rotation with 
$\alpha= \pi/2$. The solutions of these Bogomol'nyi equations
are the $1/2$-BPS states described above.

 The system allows a new type of  static Bogomol'nyi equations 
given by  
\begin{eqnarray} 
\bB =\nabla\phi, \ \  
\bD =\nabla\chi\,,
\label{quarterbogo} 
\end{eqnarray}
with the Gauss law constraint $\nabla\cdot\bD=0$,    
where $\bD$ is the momentum conjugate to ${\bf A}$ defined by
$D_i\equiv \partial {\cal L}/\partial\dot{A}_i$.
The derivation of these equations is relegated to Appendix. 
From the supersymmetric view point, the first equation in 
(\ref{quarterbogo}) breaks half of the sixteen supersymmetries 
of D3-branes
and the second  breaks further half of 
the remaining eight supersymmetries.
The related BPS configurations with $\bB,\bD\neq 0$ leave 
only four supersymmetries unbroken and, hence,
they are so-called
$1/4$-BPS states\cite{bergman}.

To show that the static solutions of the Bogomol'nyi equations 
do satisfy the original field equations, let us expand the 
Lagrangian
density at the background that satisfies the Bogomol'nyi 
equations. With the Bogomol'nyi equations  used for the 
background, one may easily
 find that the Lagrangian ${\cal L}$ can be expanded as
\begin{eqnarray} 
{\cal L}=-{\bar\bB}^2-\bigr(\bar\bB\cdot \delta \bB
- \bar\bE \cdot \delta\bE+
\nabla\bar\phi\cdot \nabla\delta\phi+
\nabla\bar\chi\cdot \nabla\delta\chi\bigl)+O(\delta^2)\,,
\label{expansion} 
\end{eqnarray}
where the quantities with  a bar represent the  solution 
of the Bogomol'nyi equations and we have used the fact that 
$\bar{\bf E}=\bar\bD$. When the Bogomol'nyi
equations are further used, the terms in the parenthesis
in (\ref{expansion}) become  purely surface terms and, 
consequently, the action is minimized for an arbitrary variation of 
the fields. This verifies that the background indeed satisfies 
the original field equations. A $1/4$-BPS configuration can be found 
 explicitly by solving  the 
Bogomol'nyi equations:
\begin{eqnarray}  
&&\bD =-\nabla {q\over 4\pi |\br-\bx|}\,,\ \ 
\chi=-{q\over 4\pi |\br-\bx|}\,,\ \ \nonumber\\
&&\bB=
-\nabla {g\over 4\pi |\br-\by|}\,,\ \ 
\phi=-{g\over 4\pi |\br-\by|}\,,
\label{multisolutionaa} 
\end{eqnarray}    
where $\bx$ and $\by$, respectively, denote the positions of 
the F-string and the D-string. 
To find the shape of the solution, we first note that $Y^I$ is given by
${\hat{e}}_1^I \phi+{\hat{e}}_2^I \chi$. Hence the spike of 
the monopole is directed to ${\hat{e}}_1^I$, while that of 
 the fundamental
string  to the direction ${\hat{e}}_2^I $.  Particularly, 
they are perpendicular to each other in the transverse space..
The configuration of the above $1/4$-BPS state is illustrated in Fig. 2.

\begin{figure}
\epsfxsize=3.5in
\centerline{
\epsffile{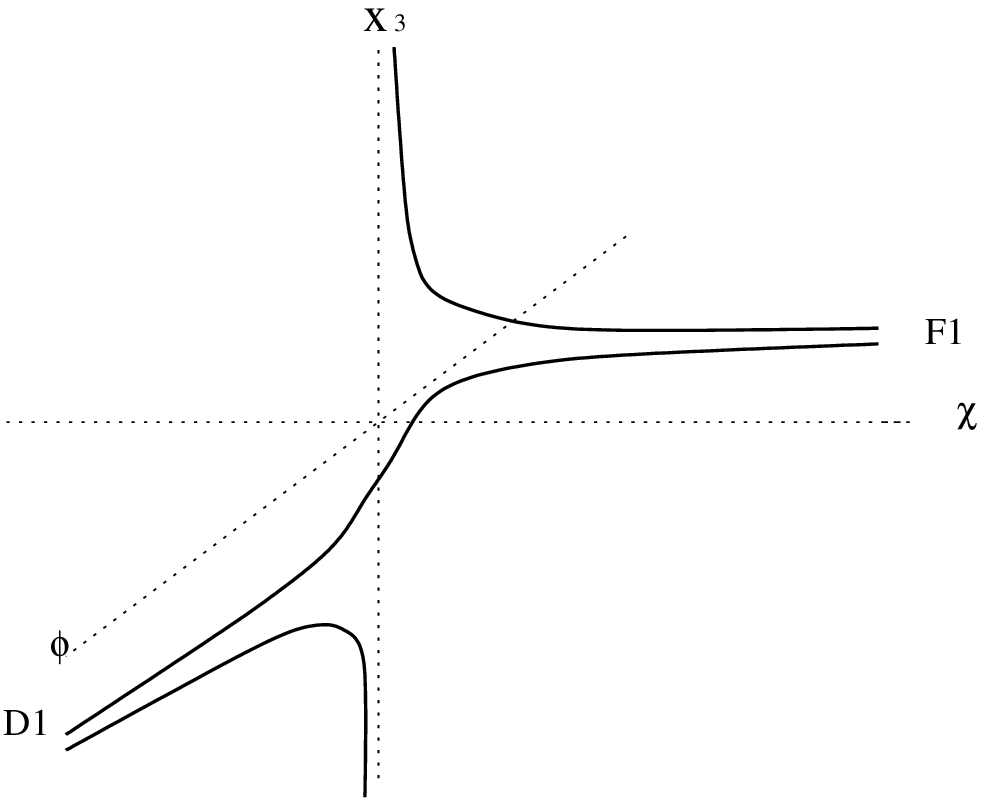}
}
\vspace{.1in}
{\small Figure~2:~ The spikes of two separated F-string and D-string
are  perpendicular to each other in the transverse space. 
This corresponds to a $1/4$-BPS states.
}
\end{figure}


 Due to the BPS nature of the 
configuration, 
the static forces between the strings are again  vanishing.  
Electromagnetic static force does not appear since one is charged 
electrically and the other carries only magnetic charge.  Hence,
the static scalar force  should  not exist. This can be understood as 
follows. The F-string possesses  the scalar charge component that  
produces a nonvanishing gradient of $\chi$-field. But the 
D-string cannot feel the $\chi$-field since it carries $\phi$-charge
only.  Consequently, the scalar interaction does not appear in this 
case, and this fact can be stated more effectively by saying that
 the two scalar charges have  different  flavor 
indices.

Let us imagine the configuration where two $1/2$-BPS  dyons
located at different positions are pulled out in two 
arbitrary transverse directions.
In this case, the static force is not balanced in general,
so the corresponding configuration is not a BPS state.
However, when the transverse angle between the spikes 
takes on a special value determined by  scalar charges,
the static force may disappear.  
Indeed, the most general $1/4$ BPS solutions are given by
\begin{eqnarray}  
&&\bD =-\nabla \sum_n {q_n\over 4\pi |\br-\bx_n|}\,,\ \ 
\chi=-\sum_n{q_n\over 4\pi |\br-\bx_n|}\,,\ \ \nonumber\\
&&\bB=
-\nabla \sum_n{g_n\over 4\pi |\br-\bx_n|}\,,\ \ 
\phi=-\sum_n{g_n\over 4\pi |\br-\bx_n|}\,,
\label{multisolutionbb} 
\end{eqnarray}    
where the number of dyons and their positions are arbitrary.
In this configuration, each spike of the n-th   $(q_n,g_n)$ dyon is  
in the transverse direction 
${-g_n\over \sqrt{q_n^2+g_n^2}}{\hat{e}_1^I}
+{-q_n\over \sqrt{q_n^2+g_n^2}}{\hat{e}_2^I}$.
Namely, the directions of the spikes are not arbitrary 
because they are determined by their electromagnetic charges. 
 The 
static forces of  the configuration  are again  balanced, which 
will be analyzed in detail later on.

So far we have considered the static properties of $1/2$-BPS states
and $1/4$-BPS state. More general configurations 
of  D-strings, F-strings, or dyons are allowed in the theory, but 
they are, in general, not the BPS states. Since the static force exist 
in case of the generic non-BPS states, the corresponding 
configurations cannot be static. 
We shall exploit 
related  dynamic 
aspects of F/D strings including the $1/4$ BPS states  
in the next section.

\nopagebreak
\setcounter{equation}{0}
\renewcommand{\theequation}{\arabic{section}.\arabic{equation}}
\section{Response of  a BPS State to Fields on a Brane}
\nopagebreak
\medskip
\nopagebreak

The F/D-strings dealt in the previous section, are solitonic 
configurations of fields, 
and they are characterized by their positions,  charges 
and  mass.  Because they have  charges, the configurations  
are expected 
to respond to asymptotic excitations on the D3-brane.
In general,
these kinds of classical dynamics must be totally governed 
by 
the original field equations. However, when
the asymptotic fields on the D3-brane are weak enough,
the response of the string will be 
linear in 
the asymptotic fields, and governed by the linearized
field equations in the background configuration of a single
F/D string. 
Furthermore, in this weak field limit,
 the shape deformation 
of the well-separated F/D strings, will be  negligible and 
each string will be
moving collectively as an independent entity; 
this motion may be characterized by the time 
dependence of the center position of each  F/D-string.

To analyze the weak-field response, 
let us expand the Lagrangian density (\ref{lag})
to the quadratic order
of the small variations from a dyon solution. We choose the scalar field
$\phi= Y^I {\hat{e}}^I$ as before where ${\hat{e}}^I$ is the
direction of the scalar for the  dyon. 
The remaining components of the scalar
can also be fluctuating but it can be easily  shown that they decouple
from the dynamic modes of the dyon.   Then, the resulting 
 expression reads
\begin{eqnarray} 
{\cal L}\!=\!-\!{\bar\bB}^2\!+\!{1\over 2}\Bigl( \delta\dot\phi^2 \!
+\!(\delta\bE)^2\!-\!
(\nabla\delta\phi)^2\!-\!(\delta\bB)^2\Bigr)
\!+\! 
{Z(r)\over 2}\bigr(\cos\xi \delta\bB \!+\!\sin\xi 
\delta\bE\! -\!\nabla\delta\phi
\bigl)^2\!+\!O(\delta^3)\,,
\label{lagpert} 
\end{eqnarray} 
where
\begin{eqnarray} 
Z(r)={(\nabla\bar\phi)^2\over 1+{\bar\bB}^2 }=
{\bar\bB^2\over \cos^2\xi(1+{\bar\bB}^2) }\,,
\label{zfactor} 
\end{eqnarray}
and the quantities with a bar denote the background dyon solution
satisfying the Bogomol'nyi equations (\ref{bogomolnyi}).
The corresponding linearized Euler-Lagrange equations
can be found to be
\begin{eqnarray} 
&& {\partial^2\over \partial t^2}\delta\phi 
-\nabla^2 \delta\phi - \nabla\cdot Z(r)\delta \bG=0\,,
\label{lineara}\\
&& {\partial\over \partial t}\delta\bE 
-\nabla\times \delta\bB +\cos\xi \nabla\times Z(r) \delta\bG 
+\sin\xi{\partial\over \partial t}Z(r)\delta \bG=0\,,
\label{linearb}
\end{eqnarray} 
with the Gauss law constraint 
\begin{eqnarray} 
 \nabla\cdot\delta\bE +\sin\xi \nabla \cdot Z(r)\delta\bG=0\,,
\label{lineareq2} 
\end{eqnarray}
where $\delta\bG$ is the combination, 
$\cos\xi\,\delta\bB+\sin\xi\,\delta\bE- \nabla\delta\phi$.
For later convenience, one may rearrange the 
above set of equations as 
follows. When Eq. (\ref{lineareq2}) multiplied by $\sin\xi$ is 
added to
 Eq. (\ref{lineara}), one obtains
\begin{eqnarray} 
{\partial^2\over \partial t^2}\delta\phi 
+\nabla\cdot W^{-1}(r)\delta \bG=0\,,
\label{linearaa}
\end{eqnarray}    
where we have introduced a new function $W(r)\equiv 1+\bar{\bB}^2$.
Similarly, the remaining equations, 
(\ref{linearb}) and (\ref{lineareq2}), 
can be written as 
\begin{eqnarray} 
&& {\partial\over \partial t}(\delta\bE -\sin\xi \nabla\delta\phi)
-\cos\xi \nabla\times W^{-1}(r) \delta\bG 
-\sin\xi{\partial\over \partial t}W^{-1}(r)\delta \bG=0\,,
\label{linearbb}
\\
&&\nabla\cdot(\delta\bE - \sin\xi\, \nabla \delta\phi)
-\sin\xi \nabla \cdot W^{-1}(r)\delta\bG=0\,,
\label{linearcc} 
\end{eqnarray} 
   
Let us first consider the case where the dyon accelerates 
constantly in  weak asymptotically uniform fields. 
To find such a solution, we adopt the following ansatz:
\begin{eqnarray} 
&& \delta {\bf A}=-{t^2\over 2}\ba\cdot \nabla 
\bar{\bf A} -t \ba \bar{A}_0+\delta\tilde{\bf A}\,,
\label{ansatza}
\\
&& \delta {A}_0=-{t^2\over 2}\ba\cdot \nabla 
\bar{A}_0 -t \ba\cdot \bar{\bf A}+\delta\tilde{A_0}\,,
\label{ansatzb}\\
&& \delta {\phi}=-{t^2\over 2}\ba\cdot \nabla 
\bar{\phi} +\delta\tilde{\phi}\,,
\label{ansatzc}
\end{eqnarray} 
where again the quantities with a bar refer to the dyonic 
background solution. The first terms in 
(\ref{ansatza})-(\ref{ansatzc}) are responsible for the bulk
motion of the dyon while the second terms in   
(\ref{ansatza}) and (\ref{ansatzb})  
are the terms generated by the instantaneous Lorentz boost with 
the 
velocity $t\ba$. Finally, the third terms are assumed to be
time independent perturbations, which will be responsible 
for the remaining dynamical aspects of the dyon.
When this ansatz is inserted into the equations, 
(\ref{linearaa})-(\ref{linearcc}), one  finds that they are 
reduced to
\begin{eqnarray} 
&& \nabla\cdot [W^{-1}\delta\tilde{\bG} -\ba \bar\phi]=0\,, \ 
\nabla\times [W^{-1}\delta\tilde{\bG} -\ba \bar\phi]=0\,,
\label{reduced}
\\
&& \nabla\cdot (\delta\tilde\bE- \sin\xi\,\nabla\tilde\phi)=0\,,
\label{modifiedgauss}
\end{eqnarray} 
with  $\delta\tilde\bG$ being 
$\cos\xi\,\delta\tilde\bB+
\sin\xi\,(\delta\tilde\bE+\ba\bar{A}_0)- 
\nabla\delta\tilde\phi$.
The most general solutions of these equations are 
provided if one solves 
the following equation
\begin{eqnarray} 
(1+\bar\bB^2)(\bar\phi+S_0)\ba =
\cos\xi\,\delta\tilde\bB+\sin\xi\,(\delta\tilde\bE+\ba\bar{A}_0)- 
\nabla\delta\tilde\phi= \delta\tilde{\bG}\,,
\label{modified}
\end{eqnarray} 
together with the Gauss law (\ref{modifiedgauss}),
where $S_0$ is the integration constant and we have written the
function $W(r)$ explicitly.
When viewed from $r=\infty$, the  equation (\ref{modified})
implies the force law,
\begin{eqnarray} 
 M(\epsilon)\ba =g\,\bB_0+q\, \bE_0- 
q_s {\hat{e}}^I\nabla Y^I_0\,,
\label{forcelaw}
\end{eqnarray} 
where we have chosen the constant $S_0$ as $l(\epsilon)$ and 
($\bB_0$,$\bE_0$,$\nabla Y^I_0$) 
refer respectively to the asymptotic values 
of magnetic field, electric field, and the scalar fields.  
When  the Lorentz symmetry  of the system is used, the force law 
can be transformed into a covariant form:
\begin{eqnarray} 
 {d\over dt}\Bigl({[M(\epsilon)\!+\!q_s {{\hat{e}}^I}Y_0^I]
\,\bv\over \sqrt{1-\bv^2}}\Bigr) \!=\!g(\bB_0\!-\!\bv \times \bE_0)+
q(\bE_0\!+\! \bv \times \bB_0)-
q_s {\hat{e}}^I\nabla Y^I_0\sqrt{1\!-\!\bv^2}\,
\label{covariant}
\end{eqnarray} 
where $\bx$ and $\bv$ are, respectively, 
the position and velocity of the dyon center\cite{bak3}. The appearance of the factor,
$M(\epsilon)\!+\!q_s {{\hat{e}}^I}Y_0^I$, on the left 
side can be understood from the fact that the change in the asymptotic
value of the scalar effectively contributes to the mass by the 
corresponding length change of the F/D-string. 

For general angle $\xi$, the equations, (\ref{modified}) and 
(\ref{modifiedgauss}) can  be solved  and the 
duality symmetry upon the interchange\cite{bengtsson,parra},
\begin{eqnarray} 
\bE\  \rightarrow\  \bB\,,\ \ \  \bB\  \rightarrow\  -\bE\,, 
\ \ \ q\ \leftrightarrow\ g\,, 
\label{duality}
\end{eqnarray}
may be explicitly verified. 
For simplicity, 
let us focus on the case of 
D-string (i.e. $\xi=0,\pi$).  The solution for 
the pure D-string reads 
\begin{eqnarray} 
\delta\tilde\phi=\hat{r}\cdot\ba \varphi(r)+\br\cdot{\bf C}\,,\ 
\delta\tilde{\bf A}=\hat{r}\times\ba \varphi(r)-
{1\over 2}[\br\times{\bf C}+l(\epsilon)
\br\times{\bf a}]\,, \ \delta\tilde{A}_0=\br\cdot{\bf E_0}
\label{accsol}
\end{eqnarray}         
with the function $\varphi$ being
\begin{eqnarray} 
\varphi={g\over 8\pi}\Bigl({1} -
{[{g/ (4\pi r)}-l(\epsilon)]^2\over r^2}
 \Bigr)\,.
\label{varphi}
\end{eqnarray}
This solution involves seven integration constants.
Six of them  are given by arbitrary constant vectors, ${\bf C}$ and 
${\bf E}_0$.
The remaining one involved with the function $\varphi$ has been
fixed by the requirement that $\delta\tilde\phi(\epsilon)=o(\epsilon)$
as $\epsilon$ goes to zero. The requirement 
implies that the D-string 
is moving without any significant deformation
around 
$r=|q_s|\epsilon$ as $\epsilon$ approaches zero.
This boundary condition at $r=|q_s|\epsilon$ deserves further discussion.
If we suppose that the string is extended to the region 
$r<|q_s|\epsilon$, the boundary condition implies that 
the segment of the string in the region $r<|q_s|\epsilon$
should always move with the same velocity as the 
string segment of $r\ge |q_s|\epsilon$.
This implies physically that the asymptotic fields are 
responsible only for the motion of the latter 
string  segment ($r\ge |q_s|\epsilon$), since the movement
of the former part is controlled by hand.
Namely, the energy transfer
from the asymptotic fields to the string should result in 
the energy gain of the string segment in $r\ge |q_s|\epsilon$, 
as will be illustrated explicitly, in the next section, by 
giving the effective theory of the F/D-string dynamics.
This, in turn, means that the energy transfer 
across the boundary at
$r=|q_s|\epsilon$ should be vanishing. When there exist  
many strings, the boundary conditions on each string at 
$|\br-\bx_n|=|(q_s)_n|\epsilon$ may be specified in
the same way, and 
then the resulting dynamics of the strings will be consistent 
with
the conservation of energy and the unitarity requirement.
(Note that our regularization is such that the transverse lengths 
of all strings are  the same.)
This will be further discussed in Sec. V in the context of 
many D3 branes where strings of finite length are 
naturally realized  as connecting two 
 D3-branes.

The accelerated charges should emit electromagnetic and 
as well as the scalar radiations. Using explicit solution, one 
finds that the radiated  power densities   
for the electromagnetic and the scalar fields  
are, respectively, given by
\begin{eqnarray} 
T^I_{\rm scalar}= {\hat{e}}^I {q_s^2(\ba \cdot \hat{R})^2\over 
16\pi^2 R^2}\,,\ \ \ \ 
T_{\rm em}= 
{(q^2+g^2)(\ba \times \hat{R})^2\over16\pi^2 R^2} 
\label{radiation}
\end{eqnarray}
where vector, ${\bf R}$, denotes the difference between the 
observational point and the position of F/D-string at retarded
time. (A detailed discussions  can be found in Ref. \cite{bak1,bak2}
where  an accelerated dyon in $SU(2)$-Yang-Mills-Higgs theory is 
dealt with.)

Let us now consider the motion of F/D-string due to
 more general asymptotic 
excitations. For our purpose, consideration of  the 
 harmonic 
time dependence $e^{-i\omega t}$  is sufficient
 since 
the equations are linear.
Eqs. (\ref{linearaa}) and (\ref{linearbb}), then,  become
\begin{eqnarray} 
&&\delta\phi={1\over \omega^2} \nabla\cdot {\bb}
\label{phisol}\\
&&\delta{\bf A}=-{1\over \omega^2}\Bigl[ \cos\xi\, 
\nabla\times\bb +i\sin\xi(\omega\bb-\omega^{-1}{\nabla\nabla\cdot\bb})
-i\omega \nabla\delta A_0\Bigr]\,,
\label{vectorsol}
\end{eqnarray} 
with 
\begin{eqnarray} 
\bb=W^{-1}(r)\Bigl[
\cos\xi\,\delta\bB+\sin\xi\,\delta\bE- 
\nabla\delta\phi\Bigr]\,.
\label{excitation}
\end{eqnarray}
There is no equation that constrains $\delta A_0$, which reflects 
the fact that
$\delta A_0$ is a pure gauge degree of freedom.   
Inserting (\ref{phisol})
and (\ref{vectorsol}) into the right side of (\ref{excitation}), 
we obtain 
the equation for $\bb$ that reads
\begin{eqnarray} 
(\nabla^2+\omega^2)\bb+ \omega^2(\nabla\bar\phi)^2 \bb=0  \,,
\label{harmoniceq}
\end{eqnarray} 
where $(\nabla\bar\phi)^2= q_s^2/(16\pi^2r^4)$ for a dyon. 
This equation is precisely the one dealt in Ref. \cite{callan} 
for the S-matrix analysis of D-string. The scattering analysis in 
Ref. \cite{callan,gubser} shows that the low frequency wave from 
the asymptotic 
region does not pass through the throat of the F/D-string. This is 
the case where the Dirichlet boundary 
condition\cite{polchinski} of the string 
attachment to the D3-brane
in the Type IIB string picture, is indeed realized.  On 
the other hand, if the high frequency wave is incident upon the F/D-string,
the wave freely travel through the throat of the F/D-string. 
Hence at high  frequency or energy, the theory requires  
boundary data on the energy flux. For this reason, we shall restrict 
our discussion below to low enough energy dynamics, 
where the boundary 
flux is negligible.

\setcounter{equation}{0}
\renewcommand{\theequation}{\arabic{section}.\arabic{equation}}
\section{Interactions between  BPS States}
\nopagebreak
\medskip
\nopagebreak

As discussed previously, the Dirac-Born-Infeld 
theory allows
multi-F/D-string configurations. These F/D-strings 
may interact via electromagnetism and  
the scalar field on 
the D3-brane. The static multi-string configuration may or 
may not be possible depending on whether 
their static potential vanishes or not. Since any BPS
state that  satisfies the Bogomol'nyi 
equations is static,
the 
electromagnetic force 
should 
be exactly canceled by the scalar 
force. As will be shown below, this 
cancellation occurs 
if  the charges carried by the strings 
satisfy  certain 
conditions. If this  condition does not 
hold for certain
configurations, they cannot be static 
due to the existence
of the static potential between strings.
In this Section, we 
shall exploit the low-energy interactions between well separated 
F/D-strings,
 based on the dynamics of a single F/D 
strings discussed 
in the previous section.

Let us begin by the discussion on the free 
kinetic terms 
for the objects involved with many-string dynamics. 
The dynamical degrees of freedom for the effective 
theory will
be  asymptotic electromagnetic fields, 
six scalar fields and the center positions of 
F/D-strings.
The kinetic terms of the asymptotic electromagnetic 
fields or the scalar fields can be easily obtained 
from the original Dirac-Born-Infeld Lagrangian. 
Since we are 
interested  in  low-energy dynamics, the linearized
version of the Dirac-Born-Infeld Lagrangian will be enough
to describe the excitations on the
D3-branes. 
The kinetic terms for each F/D-string are also
easily identified  using 
the Lorentz symmetry of the system. 
The energy of a single F/D-string will 
become  as $M(\epsilon)/\sqrt{1-\bv^2}$ when the static 
configuration is Lorentz-boosted with a velocity $\bv$.
This represents  a moving F/D-string with a constant 
velocity. The kinetic term for this energy is evidently
provided by $-M(\epsilon)\sqrt{1-\bv^2}$, which will serve
as the free Lagrangian for  F/D-string.

Let us turn 
to the    
interaction terms for the low-energy 
effective  theory. 
The information on the 
interaction vertices is coded in the equations of motion
obtained in the last Section.
Namely, the interaction terms determine the forces in the 
F/D-string motion. They also serve as a source  for 
the asymptotic fields.    

The above discussion  is summarized in the following 
action,
\begin{eqnarray}
\label{effaction}
I_{\rm eff}\!=\!\int d^4x\{{1\over 4} F^{\mu\nu}
F_{\mu\nu}
\!-\!{1\over 2}F^{\mu\nu}
(\partial_\mu A_{\nu}\! -\!\partial_\nu A_{\mu})
\!-\!{1\over 2} \partial_\mu Y^{I} 
\partial^\mu Y^{I}\} \!-\!
\!\int\! dt \sum_{n} M_n(\epsilon)
\sqrt{1\!-\!{\dot \bx}_n^2}
\!+\!I_{\rm int}
\end{eqnarray}
with
\begin{eqnarray}
I_{\rm int}\!=\!\int dt \Bigl[ -\sum_{n} q_{sn} 
{\hat{e}}_n^{I} Y^{I}\sqrt{1\!-\!{\dot \bx}_n^2}
- q_n (A^{0}\!-\!{\dot\bx}_n\cdot 
{\bf A})  -g_n (
C^{0}-{\dot\bx}_n\cdot {\bf C})\Bigr]
\end{eqnarray} 
where $C_{\mu}(x)$, as  a function of $F_{\mu\nu}$,
is defined as
\begin{eqnarray} 
\label{cmusol}
C^{\mu}(x)=-\int d^4 x' (n\cdot \partial)^{-1}(x,x')
n_\nu\,^*\! F^{\mu\nu}(x')\,,\ \  (
^*\! F^{\mu\nu}={1\over 2}\epsilon^{\mu\nu\lambda\delta} 
F_{\lambda\delta})\,.
\end{eqnarray} 
Here, $n^\mu$ may be any fixed, spacelike, unit vector, 
and  Green's 
function $(n\cdot \partial)^{-1}$ is realized by
\begin{eqnarray} 
\label{kernel}
(n\cdot \partial)^{-1}(x,x')&=&{1\over 2}\int^\infty_0 d\xi[ 
\delta^4(x-x'-n\xi)
- \delta^4(x-x'+n\xi)]\,.
\end{eqnarray} 
To be definite, we shall choose the location of the 
symmetric infinite 
Dirac string by $n^\mu=(0, \hat{n})$. 
The magnetic interaction  terms given by the function 
$C_\mu (x)$ are borrowed 
from Schwinger's Lagrangian  formulation 
for both electric  and magnetic 
charges\cite{schwinger}. It is straightforward to 
verify that the equation of motion for an F/D-string 
(\ref{covariant}) can be derived as the Euler-Lagrange 
equation from the above Lagrangian. Furthermore, the effective 
action describes all  the low-energy processes
involving electromagnetic and scalar excitations arising 
from the motion of the F/D-string. These include radiations 
from the accelerated F/D-strings, which may be shown 
to agree with the expression (\ref{radiation}). 
Especially, the low-energy interactions between F/D-strings 
are understood as the mediation of the asymptotic fields
of D3-branes.  For well separated, slowly moving  
F/D-strings, these interaction can be clearly seen 
by integrating out  the asymptotic fields. 
This is achieved by solving the 
field equations for the asymptotic fields and by inserting 
the solution into the action (\ref{effaction})[see
Ref. \cite{bak3} for detailed procedure.]. 
The resulting effective Lagrangian to the quadratic
 order in  velocities,  is given by
\begin{eqnarray}
\label{dyonaction}
L_D\!&=&\!\!\!\!-\!\sum_{n} M_n(\epsilon)\!+\!
{1\over 2}\sum_{n} M_n {\dot \bx}^2_n \! -\! 
{1\over 16\pi}\sum_{n\neq m}(q_s)_n(q_s)_m {\hat{e}}_n^I 
{\hat{e}}_m^I
{ |\dot\bx_{nm}|^2 \over |\bx_{nm}|}\nonumber\\
\!\!\!\!&-&\!\!\!\! {1\over 16\pi}\sum_{n\neq m}
(q_n g_n-g_nq_m)
\dot{\bx}_{nm}\cdot\omega 
(\bx_{nm})
\nonumber\\
\!\!\!\!&-&\!\!\!\!{1\over 16\pi}\!\!\sum_{n\neq m} \!
[(q_s)_n(q_s)_m {\hat{e}}_n^I {\hat{e}}_m^I \!-\!q_n q_m \!-\!
g_n g_m]\left\{
{{\dot \bx}_n\cdot{\dot \bx}_m\over |\bx_{nm}| }\!+\!
{\bx_{nm}\!\cdot\!{\dot \bx}_n \,\,\bx_{nm}
\!\cdot\!{\dot \bx}_m
\over |\bx_{nm}|^3 }\right\}
\nonumber\\
\!\!\!\!&+&\!\!\!\!{1\over 8\pi}\!\!\sum_{n\neq m} {
(q_s)_n(q_s)_m {\hat{e}}_n^I {\hat{e}}_m^I \!-\!q_n q_m \!-\!
g_n g_m\over
|\bx_{nm}|}\,,
\end{eqnarray}
with $\bx_{nm}\equiv \bx_n-\bx_m$, 
where $\omega^i(\br)$ denotes the unit-monopole static 
vector potential (with a symmetrically-located 
infinite string), 
\begin{eqnarray} 
\omega (\br)=-  {\hat{n}\times 
\hat{r}\over 
r-\hat{n}\cdot \br}+
{\hat{n}\times \hat{r}\over 
r+\hat{n}\cdot \br}\,.
\label{dirac}
\end{eqnarray}

The static potential term (the last term in $L_D$)
between two F/D-strings  vanishes
if their charges satisfy the condition
\begin{eqnarray}
J_{nm}\equiv \sqrt{q_n^2+g_n^2}\sqrt{q^2_m+g^2_m} 
{\hat{e}}_n^I {\hat{e}}_m^I \!-\!q_n q_m \!-\!
g_n g_m =0      
\label{condition}
\end{eqnarray}
where we have used the relation $q_s=\sqrt{q^2+g^2}$ 
between the scalar  and electromagnetic charges of a 
single F/D-string.
Namely, the strength of the static potential 
is solely determined by  $J_{nm}$.

\begin{figure}
\epsfxsize=3.5in
\centerline{
\epsffile{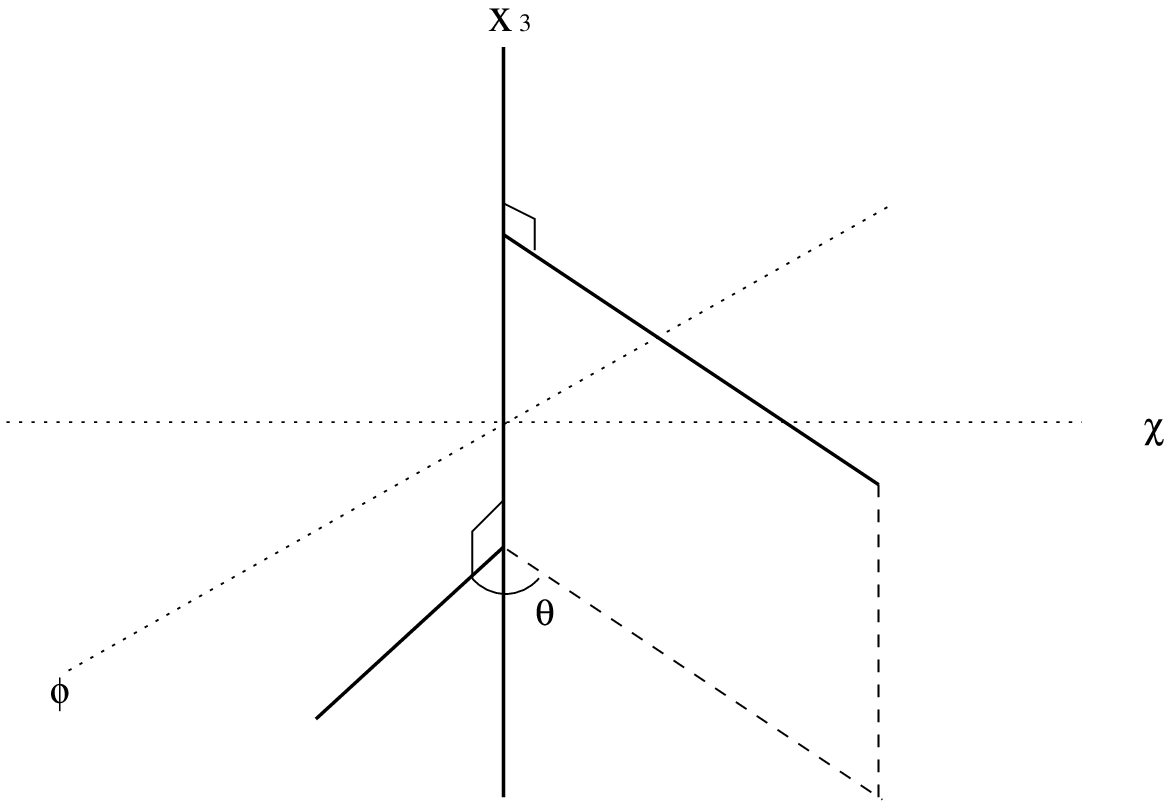}
}
\vspace{.1in}
{\small Figure~3:~The transverse angle of the spikes of two 
 separated dyonic strings  takes any values.  
There, in general,  exist  static forces  depending on 
whether they are BPS states or not. Here, the strings are schematically 
described
by lines in the transverse space.}
\end{figure}


For generic 
configuration of two 
F/D-strings  depicted in Fig. 3, the static potential
does not vanish in general since 
the transverse angle $\theta$ 
  can be arbitrary. 
But for BPS configurations discussed in Sec. I,
the static potential should, of course, be zero. 
Fig. 4 illustrates  
two cases with  vanishing  static potential.
Fig. 4a describes  two 
D-strings whose spikes are directed in the same direction of the  
transverse six-space. The magnetic force is canceled exactly by 
the attractive scalar force. The second configuration in Fig. 4, 
where  a D-string and an anti-D-string are  exactly
in opposite direction in the transverse space, also satisfies 
the condition on the charges. The repulsive force between the scalar 
charges cancels the attractive magnetic force.
We now turn to the case of $1/4$-BPS states. 
For the $1/4$-BPS state in Fig. 2, where the transverse
direction of a F-string is perpendicular to that of the D-string, 
the scalar force as well as the electromagnetic force 
 totally disappear. 
For the more general $1/4$-BPS states in 
(\ref{multisolutionbb}), we note that
%
%
\begin{eqnarray}
\cos\theta_{nm}=
{q_n q_m +
g_n g_m\over \sqrt{q_n^2+g_n^2}\sqrt{q^2_m+g^2_m}} \,,   
\label{angle} 
\end{eqnarray}
and, hence, the static potential again vanishes, i.e.  $J_{nm}=0$.
Thus, the static potential vanishes for all 
$1/2$ and  $1/4$ BPS states 
introduced in Sec. I.

\begin{figure}
\epsfxsize=5.0in
\centerline{
\epsffile{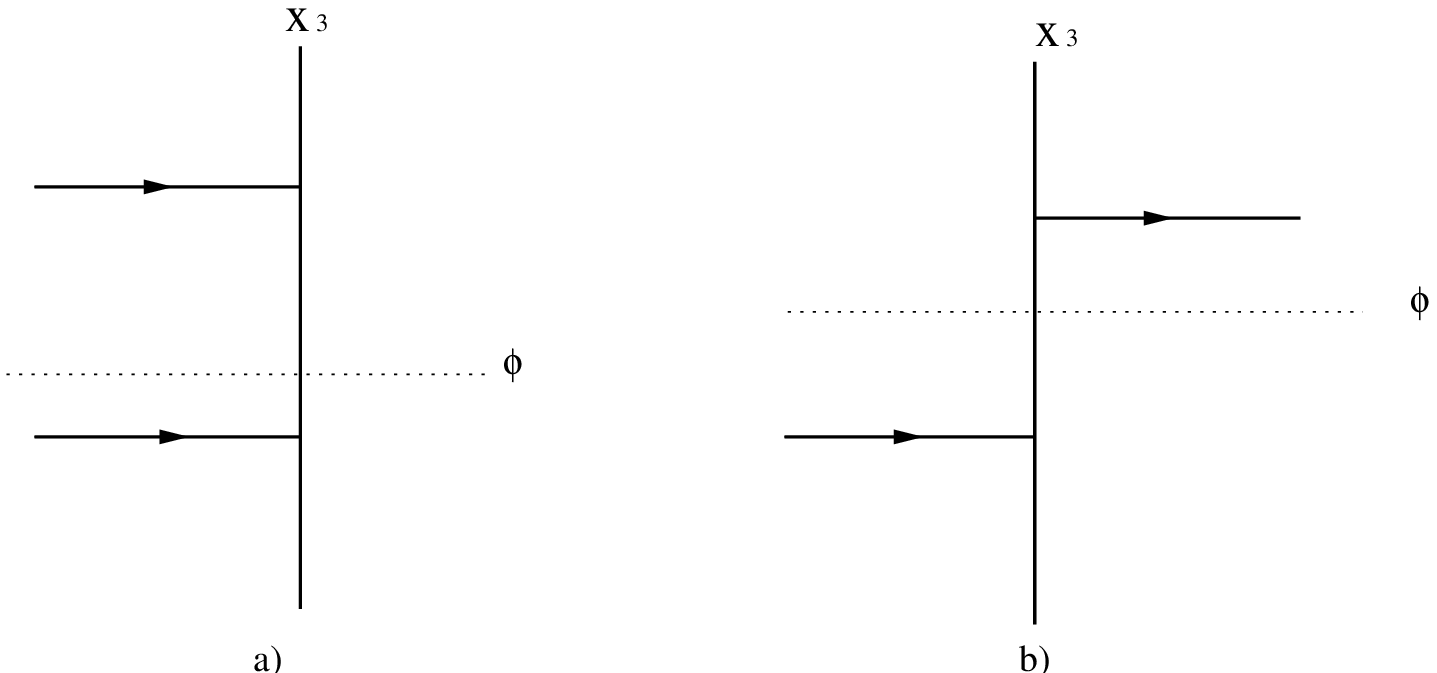}
}
\vspace{.1in}
{\small Figure~4:~The static forces disappear in cases of 
 a) parallel D-strings and b) anti-parallel D-string and 
anti-D-string. The arrow toward the brane denotes D-string, whereas
the arrow that leaves the brane is for anti-D-string.}
\end{figure}

When the BPS conditions on charges 
 are satisfied, the Lagrangian (\ref{dyonaction})
becomes purely kinetic. In other words, the classical 
trajectories are given in terms of the geodesic motions 
in the geometry defined by the kinetic term. 
For the $1/2$ BPS-states of dyons (i.e. $q_n g_m-q_mg_n=0$ 
with parallel or anti-parallel spikes for all $n$ and $m$ ), 
the Lagrangian is reduced to
\begin{eqnarray}
\label{bpsaction}
L_{1/2}=-\sum_{n} M_n(\epsilon)\!+\!
{1\over 2}\sum_{n} M_n {\dot \bx}^2_n \! -\! 
{1\over 16\pi}\sum_{n\neq m}(q_s)_n(q_s)_m \sigma_{nm}
{ |\dot\bx_{nm}|^2 \over |\bx_{nm}|}\,,
\end{eqnarray} 
where $\sigma_{mn}$ denotes the signature of ${\hat{e}}_n^I {\hat{e}}_m^I$
depending on the spikes involved are parallel or anti-parallel.
Especially, this Lagrangian  has been obtained in Ref. 
\cite{papadopoulos}, 
when only parallel D-strings ($\sigma_{nm}> 0$)are present on 
the D3-brane.

For the configuration of general $1/4$-BPS states, again the static 
potential term again drops out from the effective
Lagrangian:
\begin{eqnarray}
\label{bpsactionaa}
&&L_{1/4}=-\sum_{n} M_n(\epsilon)\!+\!
{1\over 2}\sum_{n} M_n {\dot \bx}^2_n \! -\! 
{1\over 16\pi}(g_ng_m+q_nq_m)
{ |\dot\bx_{nm}|^2 \over |\bx_{nm}|}\nonumber\\
&&\ \ \ \ \ \ \ \ \!-\!{1\over 16\pi}\sum_{n\neq m}
(q_n g_n-g_nq_m)
\dot{\bx}_{nm}\cdot\omega 
(\bx_{nm})\,.
\end{eqnarray}     
Especially, for the $1/4$-BPS configuration of Fig. 2 (i.e. F-string and 
D-string whose spikes are perpendicular),   
even the kinetic interactions disappear and the related geometry  
becomes totally flat.

\setcounter{equation}{0}
\renewcommand{\theequation}{\arabic{section}.\arabic{equation}}
\section{Conclusions}
\nopagebreak
\medskip
\nopagebreak      
 
In this note, the various BPS states  have been identified by 
solving the Bogomol'nyi equations of 
the Dirac-Born-Infeld action with six transverse 
scalar (coordinates) fields. The basic elements of the 
static BPS configurations  
 comprize electrically charged  F-string, magnetically 
charged D-string
and their dyonic bound stretched into 
the transverse six space. The more generic $1/2$ BPS  states
describe  many (anti-)parallel strings 
with their charges  restricted accordingly. We have also 
identified the generic 
$1/4$-BPS states  including  the case where the  spikes of a 
F-string and a D-string  are 
perpendicular in the transverse space.    

The asymptotic excitations on 
D3-branes consist of electromagnetism 
and the six scalar fields. We have determined the response 
of a F/D-string by obtaining the equations of 
motion  for the center position of the string. Based on this
analysis, we have constructed the low-energy effective 
field theory involving 
 many F/D-strings and the 
asymptotic fields on the D3-brane. 
When the massless degrees of freedom are integrated, the 
two-body interactions between  strings for various 
configurations are fully identified.  
   
We have found that the static potential between the 
two F/D-strings disappear  for all cases of 
the $1/2$ and the $1/4$ BPS states discussed 
in Sec. I. 
The interactions of  F/D-strings found in this note, are shown 
to be valid for low enough energies, whereas 
the description may be seriously affected for high 
enough energies.
The existence of  possible energy flow to the throat 
of a  string at high energy scattering  implies  that 
the unitariry of the theory may even break down.
But if the F/D-strings starting from a D3-brane end on another D3
brane at a finite transverse distance, the 
unitarity should  be preserved because the energy from one brane that 
passes through the throat, reappears in the other brane. 
Because  at least two 
D3-branes are involved here, the theory should be described by the 
non-Abelian Dirac-Born-Infeld action\cite{tseytlin,brecher,ferretti}. 
In this non-Abelian version, it is 
expected that the two-body interactions of F/D-strings are
determined  by the dynamics of all the branes 
where the F/D-strings end. For example, in the case of two D strings
ending commonly on the first and the second branes, the total 
interactions will be the sum 
 of those on the first and the second branes.
These indeed can be confirmed  
by using the $N=4$ $SU(N_c)$ Super-Yang-Mills theory 
as a limiting version of the full-fledged non-Abelian
Dirac-Born-Infeld actions\cite{soojong}. In this sense, we have  
obtained the interactions of F/D-strings from the view point of just one
D3-brane where the F/D-strings end commonly. 

\bigskip
\begin{center}
{\bf ACKNOWLEDGEMENTS}
\end{center}
\ \indent
The authors would like to  thank Prof. Soo-Jong Rey  for 
enlightening discussions.
\hfill
   
\bigskip
\bigskip

\vfill
\eject

\appendix
\begin{center}
{\large {\bf APPENDIX}:  
Derivation of the Bogomol'nyi Equation for $1/4$ BPS states}
\end{center}
\setcounter{section}{0}
\setcounter{equation}{0}
\renewcommand{\theequation}{\Alph{section}\arabic{equation}}
\nopagebreak
\nopagebreak
\indent

To obtain the Bogomol'nyi equations
for the scalar system with $K$ in 
(\ref{detttt}), let 
us first note 
that the determinant $K$ can be cast in the form,
\begin{eqnarray} 
 K=1+W-u^a M_{ab} u^b,.
\label{a01} 
\end{eqnarray}
where $u^i=E^i$ for $i=1,2,3$, $u^4=\dot\phi$ and 
$u^5=\dot\chi$. Since the determinant $K$ in 
(\ref{detttt}) 
is quadratic 
in $u^a$, the matrix $M_{ab}$
do not depend on  $u^a$ at all. For future use, let us give 
the explicit form of the function 
$W$ given by
\begin{eqnarray} 
 W={\bf B}^2+ (\nabla\phi)^2 +(\nabla\chi)^2
+( {\bf B}\cdot\nabla\phi)^2+({\bf B}\cdot \nabla\chi)^2 
+(\nabla\phi\times \nabla\chi)\,.
\label{a02} 
\end{eqnarray}
Using the definition $p_a\equiv {\partial {\cal L}\over \partial 
u^a}$  with 
${\cal L}=1-K^{1\over 2}$, one finds that 
the canonical momenta $p_a$ are related with $u^a$ by
\begin{eqnarray} 
p_a={M_{ab}u^b\over \sqrt{1+W-u^c M_{cd} u^d}}\,.
\label{a03} 
\end{eqnarray}
With help of the inverse 
$M^{ab}\,(M_{ac}M^{cd}=\delta^d_a)$, we find a relation
\begin{eqnarray} 
p_aM^{ab}p_b={u^a M_{ab}u^b\over 1+W-u^c M_{cd} u^d}\,,
\label{a04} 
\end{eqnarray}
Hence,  $u^a$ is 
given in terms of $p_b$ by
\begin{eqnarray} 
u^a={M^{ab}p_b\sqrt{1+W}\over \sqrt{1+p_c M^{cd} p_d}}\,.
\label{a05} 
\end{eqnarray}
The Hamiltonian, ${\cal H}\equiv 
u^a p_a -{\cal L} $ is then
\begin{eqnarray} 
{\cal H}+1=\sqrt{1+W}\sqrt{1+p_c M^{cd} p_d}={{1+W}\over 
\sqrt{1+W-u^c M_{cd} u^d}}\,.
\label{a07} 
\end{eqnarray}
The direct evaluation of the inverse $M^{ab}$ is 
too complicated, so we shall follow an  alternative route.
First, let us note that the 
static condition $u^4=u^5=0$ is required for the 
minimization of  the Hamiltonian (\ref{a07}).
Using the static condition $u^4=u^5=0$ that
implies  $M^{4b}p_b=M^{5b}p_b=0$ 
(see (\ref{a05})), the momenta $p_4$ and $p_5$
can be eliminated from   the Hamiltonian ${\cal H}$
in (\ref{a07}) with a little algebra. The resulting static
Hamiltonian reads
\begin{eqnarray} 
{\cal H}+1=\sqrt{1+W}\sqrt{1+D_i N^{ij}D_j }\,,
\label{a08} 
\end{eqnarray} 
where $N^{ij}$ is the inverse of $3\times 3$ 
matrix $M_{ij}$ 
defined by 
$M_{ij}N^{jk}=\delta^k_j$ and we have used the relation 
$D_i=p_i$ (for $i=1,2,3$) following  our earlier 
definition. 
It is relatively simple to find the inverse of the 
$3\times 3$ matrix, 
$M_{ij}$.  For example, the determinant, ${\rm Det} M_{ij}$ 
can be evaluated as
\begin{eqnarray} 
{\rm Det} M_{ij}=(1+W)\Bigl(1+ (\nabla\phi)^2 +(\nabla\chi)^2
+(\nabla\phi\times \nabla\chi)^2\Bigr)\,.
\label{a09} 
\end{eqnarray}
By the explicit evaluation of the inverse $N^{ij}$, one 
finds that our static Hamiltonian is explicitly given 
by
\begin{eqnarray} 
 ({\cal H}\!+\!1)^2&=&1+ (\nabla\phi)^2 +(\nabla\chi)^2
+(\nabla\phi\times \nabla\chi)^2  +{\bf B}^2+{\bf D}^2
\nonumber\\
&+&( {\bf B}\cdot\nabla\phi)^2+({\bf B}\cdot \nabla\chi)^2 +
( {\bf D}\cdot\nabla\phi)^2+({\bf D}\cdot \nabla\chi)^2 
(\bD\times \bB + {\Pi}\nabla\phi)^2\nonumber\\
&+&{  (\bD\!\times\! \bB)^2\!\!
+\!\!(\nabla\phi \!\times\!(\bD\!\times\! \bB))^2
\!\!+\!\!(\nabla\chi\! \times\!(\bD\!\times\! \bB))^2
\!\!+\!\!(\nabla\phi \!\times\!\nabla\chi
\cdot\bD\!\times\! \bB)^2
\over 1+ (\nabla\phi)^2 +(\nabla\chi)^2
+(\nabla\phi\!\times\! \nabla\chi)^2 }  \,.
\label{a10} 
\end{eqnarray}
Note  that this Hamiltonian with vanishing $\chi$ 
does agree with the one-scalar Hamiltonian in 
(\ref{hamiltonian}) when 
the static condition, $\dot\phi=0$ 
[or equivalently $(1\!+\!(\nabla\phi)^2)\Pi=\nabla\phi\cdot\bB
\!\times\!\bD$], is used.  
Finally, the Hamiltonian (\ref{a10}) may be turned into the form,
\begin{eqnarray} 
 ({\cal H}\!+\!1)^2&=& (\bB-\nabla\phi)^2 +(\bD-\nabla\chi)^2+
 (\bD\cdot \nabla\phi-\bB\cdot\nabla\chi )^2
+(1+\bB\cdot\nabla\phi+\bD\cdot\nabla\chi)^2
\nonumber\\ 
&+& {{\bf S}^2+(\nabla\phi\cdot{\bf S})^2 +
(\nabla\chi\cdot{\bf S})^2 + 
(\nabla\phi\cdot \nabla\phi\cdot{\bf S})^2 
\over 1+ (\nabla\phi)^2 +(\nabla\chi)^2
+(\nabla\phi\times \nabla\chi)^2 } \,,
\label{a11} 
\end{eqnarray}
where ${\bf S}$ denotes the combination $\bB\times\bD -
\nabla\phi\times\nabla\chi$.
Thus the Hamiltonian is bounded by a surface 
 terms by
\begin{eqnarray} 
 {\cal H}\ge \bB\cdot\nabla\phi+\bD\cdot\nabla\chi\,,
\label{a12} 
\end{eqnarray}
where the saturation  of the bound occurs if the claimed Bogomol'nyi
equations (\ref{quarterbogo}) are satisfied with the Gauss law 
constraint $\nabla\cdot\bD=0$.

\vfill
\eject


\begin{thebibliography}{99}
\bibitem{born} M. Born and L. Infeld, Proc. Roy. Soc. 
{\bf A 144} (1935) 425.
\bibitem{dirac} P.A.M. Dirac, Proc. Roy. Soc. 
{\bf A 268} (1962) 57.
\bibitem{callan} C. Callan and J. Maldacena, 
Nucl. Phys. {\bf B 513} (1998) 198.
\bibitem{gibbons} G. W. Gibbons, 
Nucl. Phys. {\bf B514} (1998), 603; Wormholes on the World 
Volume: Born-Infeld Particles and
Dirichlet p-Branes.  
hep-th/9801106 (1998); Branes as Bions,
 hep-th/9803203 (1998);
S. Deser and  G.W. Gibbons, Class. Quant. Grav.15 (1998) L35. 
\bibitem{alexander} A. A. Chernitskii, Helv. Phys. Acta 71 (1998)
274.
\bibitem{zeid} M. Abou Zeid and C.M. Hull,
Geometric Actions for D-Branes and M-Branes, 
 hep-th/9802179 (1998) 
\bibitem{bengtsson} I. Bengtsson,  Int. J. Mod. Phys. {\bf A12} (1997)
4869. 
\bibitem{hashimoto} 
 A. Hashimoto,  Phys. Rev. {\bf D57} (1998) 6441. 
\bibitem{bergman} O. Bergman,
Three Pronged Strings and 1/4 BPS States in N=4 Super-Yang-Mills Theory,
 hep-th/9712211 (1997);   O. Bergman and B. Kol, 
String Webs and 1/4 BPS Monopoles,
hep-th/9804160 (1998). 
\bibitem{bogomolnyi} E.B. Bogomol'nyi,
Sov.J.Nucl.Phys. 24 (1976) 449.
\bibitem{gauntlett} Jerome P. Gauntlett, 
Joaquim Gomis, Paul K. Townsend, JHEP 01 (1998) 3. 
\bibitem{silva} For the supersymmetric discussion of the bound, 
see  
S. Gonorazky, C. Nunez, F.A. Schaposnik,  and G. Silva, 
  Bogomol'nyi Bounds and the Supersymmetric Born-Infeld Theory,
hep-th/9805054 
(1998).
\bibitem{parra} A. Khoudeir and Y. Parra,  
On Duality in the Born-Infeld Theory,
hep-th/9708011, 1997.
\bibitem{bak1} D. Bak and C. Lee, Nucl. Phys. {\bf B403} 
(1993) 315.
\bibitem{bak2} D. Bak, and  H. Min, 
Phys. Rev. {\bf D56} (1997) 6665.  
\bibitem{gubser} S.S. Gubser and A. Hashimoto,
Exact Absorption Probabilities for the D3-Brane,
 hep-th/9805140 (1998).
\bibitem{polchinski} J. Polchinski,
 TASI Lectures on D-Branes,  hep-th/9611050 (1996).
\bibitem{schwinger} J.  Schwinger, Phys. Rev. 
{\bf D12} (1975) 3105. 
\bibitem{bak3} D. Bak, C. Lee and  K. Lee,
  Phys. Rev. {\bf D57} (1998) 5239. 
\bibitem{papadopoulos} 
J. Gutowski and  G. Papadopoulos, The Moduli Space 
of  World Volume Brane Solitons, hep-th/9802186 (1998).
\bibitem{tseytlin} 
A.A. Tseytlin, Nucl. Phys. {\bf B501} (1997) 41. 
\bibitem{brecher} D. Brecher, BPS States of the 
Non-abelian Born-Infeld Action, hep-th/9804180 (1998). 
\bibitem{ferretti} 
Irina Ya. Aref'eva, G. Ferretti, and A.S. Koshelev, 
Taming the Nonabelian Born-Infeld Action,
hep-th/9804018 (1998).
\bibitem{soojong} D. Bak and S. Rey, {\it} in preparation.
\end{thebibliography}
\end{document}